\documentclass[9pt,twocolumn,twoside]{pnas-new}

\templatetype{pnasresearcharticle} 

\title{Critical point for Bose-Einstein condensation of excitons in graphite}

\author[a,b]{Jinhua Wang}
\author[a,b]{Pan Nie}
\author[a,b]{Xiaokang Li}
\author[a,b]{Huakun Zuo}
\author[c]{Beno\^{\i}t Fauqu\'e}
\author[a,b,1]{Zengwei Zhu}
\author[d]{Kamran Behnia}

\affil[a]{Wuhan National High Magnetic Field Center, Huazhong University of Science and Technology,  Wuhan  430074, China}
\affil[b]{ School of Physics, Huazhong University of Science and Technology,  Wuhan  430074, China}
\affil[c]{JEIP, USR 3573 CNRS, Coll\`ege de France, PSL Research University, 11, place Marcelin Berthelot, 75231 Paris Cedex 05, France}
\affil[d]{Laboratoire de Physique et d'Etude des Mat\'{e}riaux (CNRS)\\ ESPCI Paris, PSL Research University, 75005 Paris, France}

\leadauthor{Wang}

\significancestatement{Bose-Einstein condensation of excitons (pairs of electrons and holes bound by Coulomb attraction) has become a dynamic field of research in the past couple of decades. Whilst individual excitons have been observed in many systems, a collective state of condensed excitons has proved to be more elusive. In this paper, by quantifying the mass and the density of the electron-hole pairs, we identify a critical temperature of 9.2 K and a critical magnetic field of 47 T as the cradle of the Bose-Einstein condensation of excitons in graphite. Our identification radically revises the nature of the field-induced phase transition in graphite and its boundaries.}

\authorcontributions{Z.Z. and K.B. designed the research project; J.W., P.N., X.L, and H.Z. performed the reseach; J.W., B.F., Z.Z., and K.B. analyzed data; and J.W., Z.Z., and K.B. wrote the paper.}
\authordeclaration{The authors declare no competing interest.}

\correspondingauthor{\textsuperscript{1}To whom correspondence should be addressed. E-mail:zengwei.zhu@hust.edu.cn}


\begin{abstract}
An exciton is an electron-hole pair bound by attractive Coulomb interaction. Short-lived excitons have been detected by a variety of experimental probes in numerous contexts. An excitonic insulator, a collective state of such excitons, has been more elusive.  Here, thanks to Nernst measurements in pulsed magnetic fields, we show that in graphite there is a critical temperature (\textit{T} = 9.2 K)  and a critical magnetic field (\textit{B} = 47 T) for Bose-Einstein condensation  of excitons. At this critical field, hole and electron Landau sub-bands simultaneously cross the Fermi level and allow exciton formation. By quantifying the effective mass and the spatial separation of the excitons in the basal plane, we show that the degeneracy temperature of the excitonic fluid corresponds to this critical temperature. This identification would explain why the  field-induced transition observed in graphite is not a universal feature of three-dimensional electron systems pushed beyond the quantum limit.
\end{abstract}

\dates{This manuscript was compiled on \today}
\doi{\url{www.pnas.org/cgi/doi/10.1073/pnas.XXXXXXXXXX}}

\begin{document}

\maketitle
\thispagestyle{firststyle}
\ifthenelse{\boolean{shortarticle}}{\ifthenelse{\boolean{singlecolumn}}{\abscontentformatted}{\abscontent}}{}

\dropcap{A} macroscopic number of non-interacting bosons condense to a single-particle state below their degeneracy temperature~\cite{Leggett}. This phenomenon, known as the Bose-Einstein condensation (BEC) of bosons was unambiguously detected in ultra-cold atomic gases~\cite{Anderson} seven decades after its prediction~\cite{Einstein}. The critical temperature for this phase transition depends on the mass $m^*$ and density $n$ of bosonic particles \cite{Silvera}:
\begin{equation}\label{BEC}
k_{B}T_{\rm{BEC}}= 3.31 \frac{\hbar^2}{m^*}n^{2/3}
\end{equation}
In all known cases of BEC, the particles are composite bosons made of ``elementary'' fermions. This is the case of $^4$He, which becomes superfluid below 2.17 K, and dilute cold atoms, which display BEC features below 0.17 $\mu$K~\cite{Anderson}. The difference in critical temperature reflects what is expected by Eq. \ref{BEC}. Denser fluids and lighter bosons have  higher ${T_{\rm{BEC}}}$.

The possible occurrence of BEC for excitons [bosonic pairs of electrons and holes~\cite{Mott}] has become a dynamic field of research in the past couple of decades~\cite{Snoke,Eisenstein}. Individual excitons have been observed in  semiconducting heterostructures stimulated by light creating electrons and holes in equal numbers. However, such excitons are ephemeral entities. The emergence of a collective state of spontaneously created excitons was postulated in the context of semimetal-to-semiconductor transition~\cite{Jerome1967} and was dubbed an excitonic insulator (EI).

Condensation of excitons into a collective and thermodynamically stable state \cite{Kogar,Kono2017,Butov2007,Kim,Dean,Burg,Zefang,Karni} would require three conditions: 1) a sufficiently large binding energy; 2) a lifetime exceeding the thermalization time and 3) a concentration high enough to allow a detectable degeneracy temperature. An independent issue is the identification of such a state in distinction from other collective electronic states of quantum matter. In the case of bulk 1T-TiSe$_2$, scrutinizing the plasmon dispersion has revealed a signature of EI unexpected in the alternative  Peierls-driven charge density wave (CDW)~\cite{Kogar}. 
Other indirect signatures of BEC transition have been reported in two-dimensional systems, such as quantum wells~ \cite{Butov2007}, graphene~\cite{Kim,Dean,Burg} and transition metal dichalcogenides heterostructures~\cite{Zefang,Karni}.

Here, we present the case of graphite subject to strong magnetic field where the existence of a thermodynamic phase transition is established ~\cite{Tanuma1981,Yaguchi2009,FauqueBook}. We will show that a magnetic field of 47 T provides all necessary conditions for the formation of a BEC of excitons. At this field, the gap between the two penultimate Landau subbands vanishes. One of these subbands is electron-like and the other is hole-like. The combination of vanishing gap and the large density of states (DOS) at the bottom and top of the bands permits the formation of excitons. We will show that the mass and the the density of these excitons is such that they should become degenerate below a critical temperature of the order of 9.2 K. As the temperature is lowered, this collective state survives in a narrow field window, which widens with cooling at the lower end but not at the upper end of this window. The BEC scenario provides an explanation for this contrast. Increasing the field destroys the thermodynamic stability of electron-hole pairs. Decreasing it, on the other hand, leads to a reduction of the degeneracy, gradually pulling down the critical temperature. Questions which remained unanswered in the CDW scenario for this phase transition~\cite{Yoshioka,Arnold2017} find answers by this identification.

Graphite is a semi-metal with an equal density of electrons and holes ($n=p=3\times10^{18}~ \rm{cm}^{-3}$~\cite{Brandt1988}). Above the quantum limit of 7.4 T, electrons and holes are both confined to their lowest Landau levels~\cite{ZhuNernstGraphite,Schneider}, which are each split to two spin-polarized subbands. A rich phase diagram consisting of  distinct field-induced phases emerges above $\sim$20 T, which was documented by previous studies~\cite{Benoit2013,ZhuEI2017,Leboeuf2017,Zhu2019}. Our focus, here, is the identification of the peak transition temperature as the cradle of the excitonic instability.

Theoretical calculations by Takada and Goto~\cite{Takada1998, Arnold2017}, based on the SWM  model~\cite{Slonczewski1958,McClure1957} of band structure and including self-energy corrections,  predicted that the electron spin-up and the hole spin-down subbands simultaneously cross the Fermi level   at $\sim$ 53 T. The first result of the present study is to confirm such a simultaneous crossing of the two subbands occurring at a slightly different field, namely 47 T, and its coincidence with the peak critical temperature of the field-induced phase.
\begin{figure}
\includegraphics[width=8.5cm]{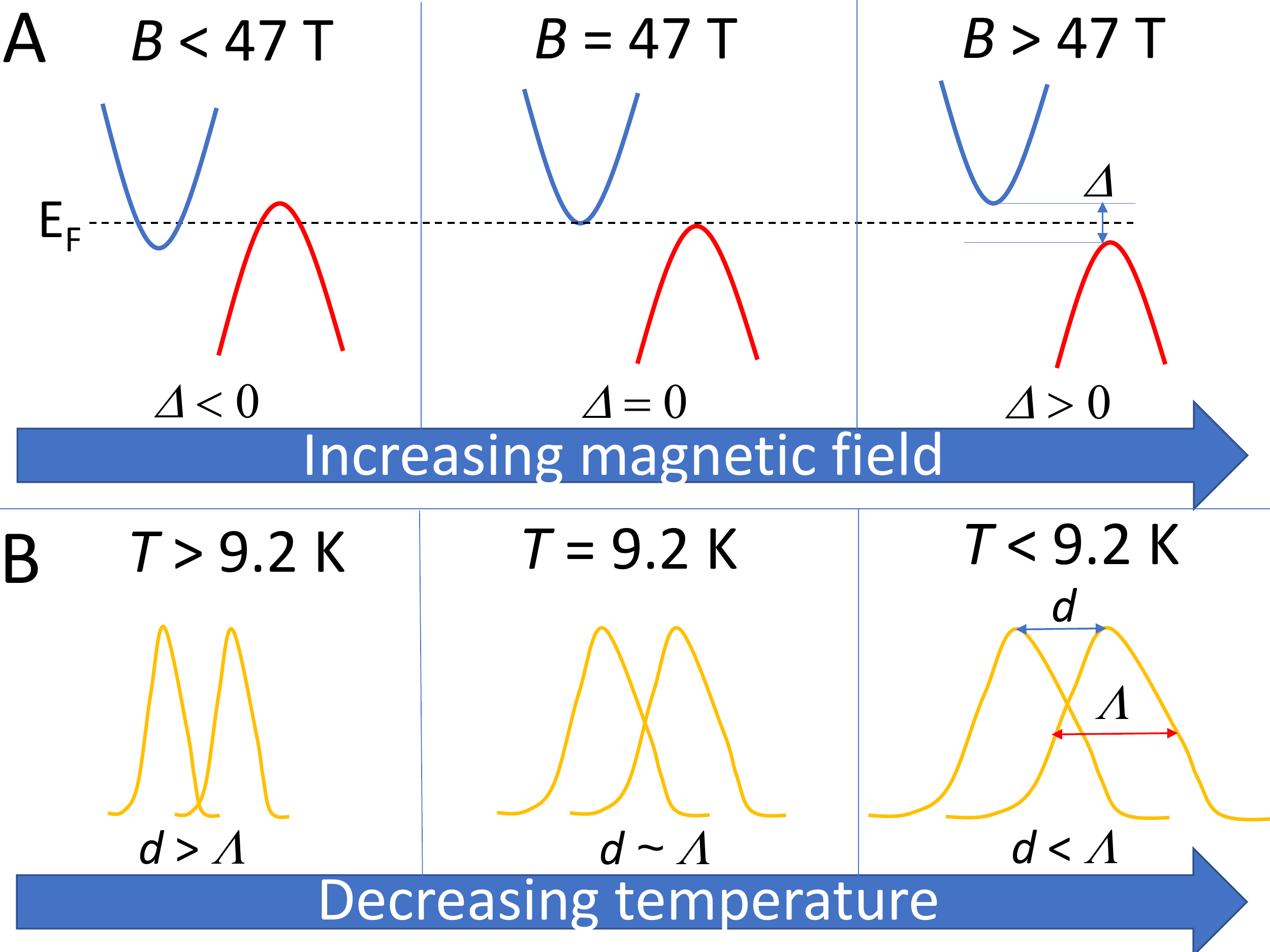}
\caption{\textbf{A critical point in (field, temperature) plane} ({\textit{A}}) As magnetic field is increased, the gap $\Delta$ between the electron spin-up and the hole spin-down subbands evolves from negative to positive. Both subbands are evacuated at 47 T and $\Delta=0$. Two other subbands with the same level index and opposite spin polarities remain occupied. ({\textit{B}}) As the temperature is lowered, the thermal de Broglie wavelength $\Lambda$ becomes longer. At 9.2 K, it  becomes comparable to the interexciton distance $d$ and BEC of exciton occurs.}
\label{CriticalPointSketch}
\end{figure}

Fig. \ref{CriticalPointSketch} presents a sketch of the proposed scenario. As the magnetic field increases, the gap between  the spin-up subband of electrons and the spin-down subband of holes changes sign. At a critical field of 47 T not only the gap vanishes but also the two subbands empty. For fields exceeding 47 T, the formation of electron-hole pairs costs a finite energy. When the field sufficiently larger than 47 T, electron-hole pairs break up. Tuning down the temperature, on the other hand, will increase the  thermal de Broglie wavelength. The BEC condensation will occur when the exciton wavelength becomes comparable to the interexciton distance,  set by carrier concentration ~\cite{Silvera}. The critical temperature of 9.2 K would be the degeneracy temperature of bosonic excitons in our picture.

We carried out a study of the Nernst effect~\cite{Behnia2016} in pulsed fields on Kish graphite samples. The Nernst effect has proved to be an extremely sensitive probe of Landau spectrum in compensated semimetals like graphite.  Quantum oscillations are most prominent in the Nernst response and dominate the nonoscillating background~\cite{Zhu2011}. However, measuring a Nernst signal in pulsed fields is challenging and a study of URu$_2$Si$_2$~\cite{ProustNernst2009} is the unique known case prior to the one presented here. Details of the experimental setup are given in the supplement~\cite{SM}.

\begin{figure}
\includegraphics[width=8.5cm]{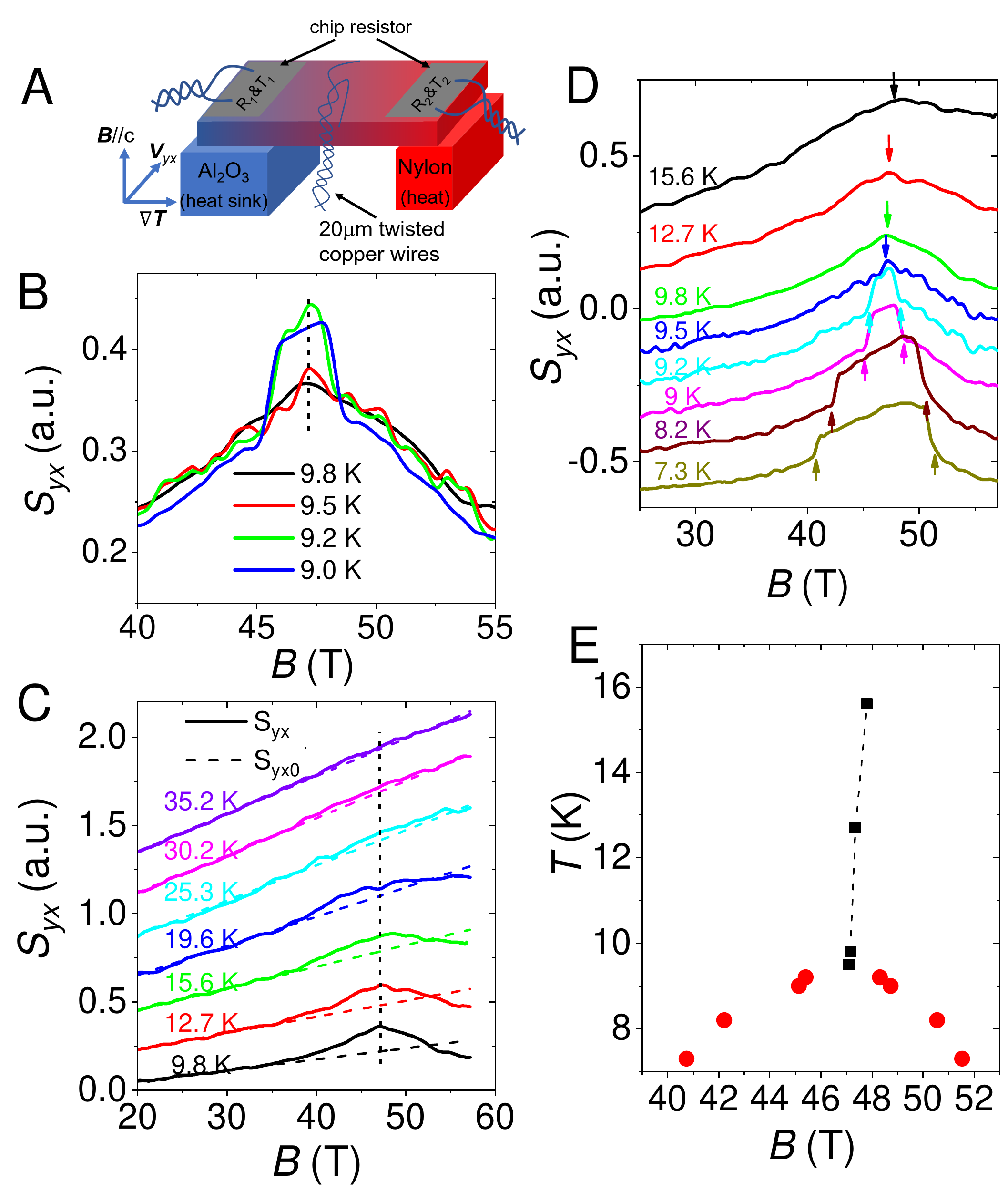}
\caption{ \textbf{Experimental signature of a critical point.} ({\textit{A}}) The sketch of the setup. ({\textit{B}}) The Nernst signal presents a structure near 47 T. The peak above  9.2 K is substituted by two distinct anomalies below. There is a jump in the Nernst signal followed by a fall as a function of increasing magnetic field. ({\textit{C}}) The broad peak centered at 47 T gradually fades away upon warming. Dashed lines are guides for the eyes. Curves are shifted for clarity. ({\textit{D}}) The evolution of the Nernst signal with warming over a broader temperature range. ({\textit{E}}) The Nernst anomalies in the (\textit{B},\textit{T}) plane  bifurcate at \textit{B}= 47 T and \textit{T}= 9.2 K. }
\label{Exp.CriticalPoint}
\end{figure}

\begin{figure}[t]
\includegraphics[width=8cm]{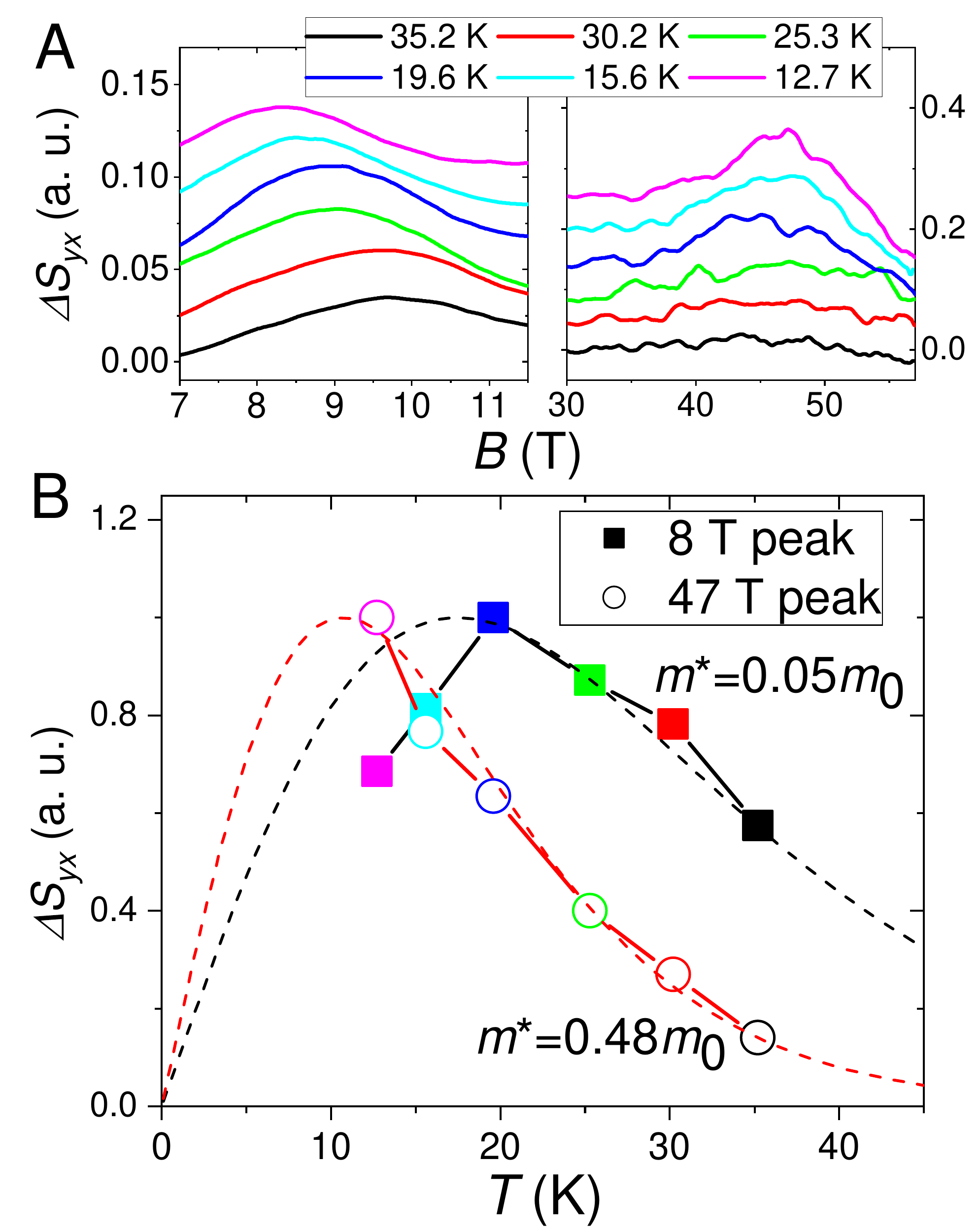}
\caption{ \textbf{Thermal evolution of the Nernst anomalies and the effective mass} (\textit{A}) The oscillatory component of the Nernst signal near quantum limit and 47 T. The low-field peak survives up to 35 K upon warming, in contrast to the high-field peak, which fades away quickly upon warming. (\textit{B}) The temperature dependence of the magnitude of the two anomalies allows to extract the effective mass and reveals a 10-fold mass enhancement.}
\label{mass}
\end{figure}
Fig. \ref{Exp.CriticalPoint} shows our Nernst data near the critical point (see the supplement~\cite{SM} for data extended up to 60 T). The signal smoothly evolves upon cooling. Below 4 K, the low-field anomaly becomes similar to what was reported in a  previous study of the Nernst effect below 45 T~\cite{Benoit2011}. Our extended data reveal additional information.

First of all, the  peak near 47 T is the only one detected up to 60 T. This indicates that the evacuation of the two Landau subbands occur simultaneously. In other words, the separation in magnetic field is too small to be detected by experiment. This interpretation is consistent with the vanishing Hall conductivity  observed near 47 T\cite{Akiba_2015}.

The second observation is that this peak suddenly splits to two distinct anomalies when $T< 9.2$ K (See Fig. \ref{Exp.CriticalPoint}{\textit{B, D, E}}). Finally, it is remarkable that the Nernst peak disappears for $T> 35$ K (See fig. \ref{Exp.CriticalPoint}{\textit{C}}). This temperature dependence allows us to quantify the high-field effective mass.

Fig. \ref{mass} compares the temperature dependence of the Nernst peaks near 8 T  and near 47 T.  What sets the thermal evolution of the amplitude of a quantum oscillation is the effective mass and the B/T ratio. The heavier the electrons, the faster the decay of the oscillating signal with warming. The larger the B/T ratio, the slower the decay. In this context, it is sriking to see that the high-field peak vanishes faster with warming than the low-field one. Quantitatively, using the Lifshitz-Kosevich formula for thermoelectric quantum oscillations, $\Omega(\rm{T})_{osc}\propto [\alpha X\coth(\alpha X)-1]/\sinh(\alpha X)$~\cite{QO2016, QO2019}, where $\alpha=2\pi^2k_B/e\hbar$ and $X=m^*T/B$, we find that the effective mass $m^*_{47\rm{T}}$= 0.48 $m_0$ and $m^*_{8\rm{T}}$ = 0.05 $m_0$  (Fig. \ref{mass}{\textit{B}}), where $m_0$ is the bare electron mass. The low-field value is consistent with the mass extracted from Shubnikov-de Haas measurements~\cite{Yaguchi2009}. Thus, there is a 10-fold field-induced enhancement in cyclotron mass of carriers. Note that recent specific heat measurement~\cite{Marcenat} find a  field-induced enhancement of the DOS.

\begin{figure}[!h]
\includegraphics[width=9cm]{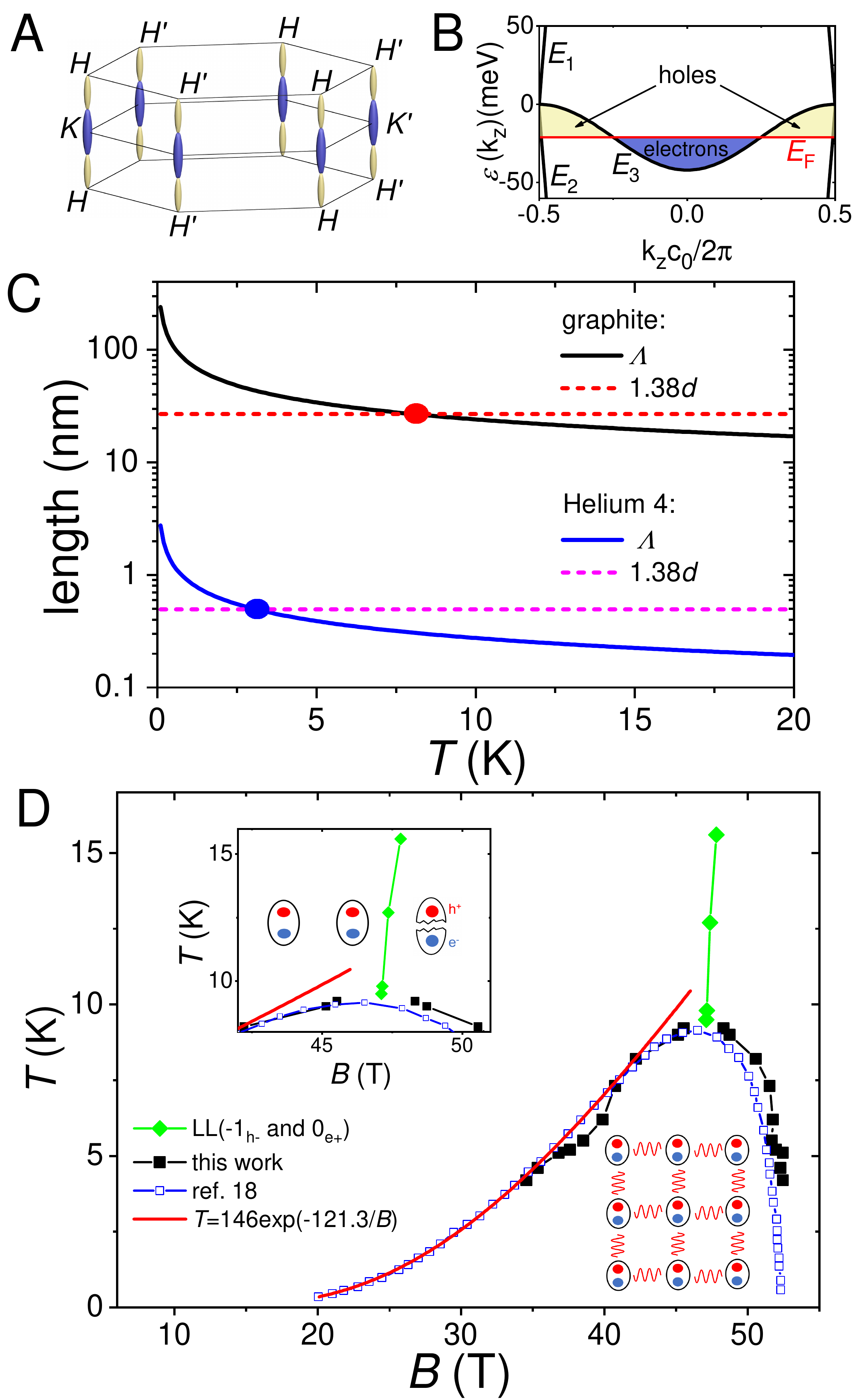}
\caption{ \textbf{Band dispersion, length scales and the boundaries of the EI phase.} (\textit{A}) Electron and hole Fermi pockets and the Brillouin zone in graphite.  (\textit{B}) Band dispersion in graphite along the $k_z$ axis. Note that electron and hole bands touch each other and extend over the thickness of the Brillouin zone along the $k_z$ axis. The system is half-filled along the $k_z$ axis and remains so in finite magnetic field. (\textit{C}) The interboson distance (times a numerical factor) and the de Broglie wavelength in liquid helium and in graphite. The two length scales become equal near the critical temperature. (\textit{D}) The phase diagram of the first field-induced phase in graphite. BEC triggers the bifurcation of the Nernst anomalies shown by green (black) symbols above (below) the critical point. At high temperature, it corresponds to the simultaneous evacuation of the (-1$_{h^{\tiny-}}$ and 0$_{e^{\tiny+}}$) Landau levels. The inset is the zoom near the critical point, where the critical temperature ceases to follow the BCS behavior and saturates to a value set by the Eq. \ref{BEC}.  The EI phase is destroyed in two different ways at its lower and upper field boundaries. Along the higher boundary, the binding energy falls below the field-induced gap. Along the lower boundary, the condensate is weakened by decreasing DOS and evolves toward a  BCS-type weak-coupling behavior.}
\label{PD}
\end{figure}
BEC occurs when the interBoson distance falls below the thermal de Broglie wavelength. In liquid $^4$He, for example, the interatomic distance is 0.358 nm and the thermal de Broglie wavelength can be estimated taking the mass of each He atom. The two length scales become equal at 5.9 K. The BEC condition which corresponds to $d_{\rm{exciton}}=\Lambda_{\rm{exciton}}/1.38$ \cite{Silvera} occurs at 3.1 K~\cite{London1938} to be compared with the superfluid critical temperature (2.17 K).   The difference can be quantitatively explained by taking to account interactions, which lead to an effective mass larger than the bare mass\cite{Vakarchuk}.

In our case, the interexciton distance and the de Broglie length are both anisotropic. According to the most extensive set of de Haas-van Alphen data ~\cite{Schneider}, the radius of Fermi surface along the c-axis is seven to nine times longer than in the basal plane. The effective mass is also very anisotropic and the  ratio of out-of-plane to in-plane masses is estimated to exceed 90~\cite{McClure1957}. In the basal plane, the two relevant length scales can be estimated unambiguously. The frequency of quantum oscillations yields the interparticle distance in the basal plane, which is 18 nm for both electrons and 21 nm for holes (Table \ref{distance}). Thus, the interexciton distance in the basal plane is $d_{\rm{exciton}}= 19.5 \pm 2~\rm{nm}$. The exciton mass would be twice the cyclotron mass resolved at 47 T, therefore, $m_{\rm{exciton}}$=0.96 $m_0$.  Using these numbers, one finds that  the BEC condition $d_{\rm{exciton}}=\Lambda_{\rm{exciton}}/1.38$ \cite{Silvera} is satisfied when $T=8$ K. This is remarkably close to the critical temperature of 9.2 K detected by our experiment. As seen in Fig. 4$C$, in graphite, the two length scales are two orders of magnitude longer than in $^4$He and in both the experimentally observed critical temperature is close to where the BEC expected.

In a compensated semimetal, charge neutrality does not impede a concomitant evolution of the density  of electrons  and holes with increasing magnetic field  across the quantum limit. This is indeed what happens in semimetallic bismuth at high magnetic fields: at 30 T, the carrier density increases to more than five times its zero field value~\cite{Zhu2017}. However, this is unlikely to happen in graphite because of its band structure\cite{Slonczewski1958,McClure1957} (see Fig. \ref{PD}$A, B$). When carriers are confined to the lowest Landau subbands, the DOS steadily increases due to the degeneracy of Landau levels. In a compensated metal, this can occur either by an enhancement in the concentration of electrons and holes, by an enhancement in mass, or a combination of both. Now, in graphite (in contrast to bismuth), electron and hole ellipsoids are aligned parallel to each other and their dispersion is similar. Moreover, and crucially, both electron and hole bands are half-filled along the $k_z$. As a result, the room for any significant modification of the Fermi wave-vector along the orientation of magnetic field and a change in carrier density is small. Thus, our analysis safely assumed that carrier density does not change between 7 T and 47 T.

\begin{table}[h]
\begin{tabular}{|c|c|c|c|c|}
\hline
  Carrier & $F_{\perp}(T)$ &$A_{\perp}(10^{12} \rm{cm}^{-2})$ &$\lambda_{\perp}(\rm{nm})$& $d_{\perp}(\rm{nm})$\\
\hline
Holes &4.7 &4.49  & 52 & 21\\
\hline
Electrons &6.45 &6.15 &45 & 18\\
\hline
\end{tabular}
\caption{The de Haas-van Alphen Effect frequencies, $F_{\perp}$ for holes and electrons in graphite with magnetic field along the c-axis~\cite{Schneider}. This allows the quantification of the areas of extremal orbit, $A_{\perp}$, the electronic wavelengths, $\lambda_{\perp}$ and the inter-particle distances, $d_{\perp}$ by using $d=\frac{\lambda_F}{\sqrt{2\pi}}$, assuming that the hole and electron Fermi surfaces are cylinders. The deviation caused by their elongated ellipsoid geometry is small. }
\label{distance}
\end{table}

The boundaries of the EI phase shown in Fig.\ref{PD}$D$ are strikingly similar to the theoretical expectations~\cite{Jerome1967,Kozlov1965}. The left (low-field) boundary evolves to a mean-field expression for critical temperature $T_{\rm{c}}(B)=T^*{\rm{exp}}(-\frac{B^*}{B})$, which is a BCS-type formula $k_{\rm{B}}T_{\rm{c}}(B)=1.14E{\rm{_Fexp}}(-\frac{1}{N(E{\rm{_F}})V})$\cite{Yaguchi2009,Iye1982} and the evolution of the DOS with magnetic field governs the evolution of the phase transition.  This expression fails as the critical point is approached, leading to the saturation of the critical temperature.
In contrast, on the right (high-field) side, the destruction of the field-induced state is abrupt and the critical temperature is pinned to a magnetic field of 53 T. This field does not correspond to the evacuation of any Landau level, as shown by the absence of any anomaly in our data. In the BEC scenario, it corresponds to the unbinding of the electron-hole pair by magnetic field (see the sketches in the Fig.\ref{PD}$D$).

 Note that only at $B=$ 47 T the critical temperature corresponds to the degeneracy temperature of excitons and the transition is,  strictly speaking, a BEC condensation . When the magnetic field exceeds 47 T, the exciton binding energy becomes lower than the band gap and the order is destroyed by unbinding. On the other hand, decreasing the magnetic field diminishes the DOS, the screened Coulomb attraction between electrons and holes and the transition occurs well below the degeneracy temperature.\color{black}

A BEC picture of field-induced phase transition would  explain its presence in graphite in contrast to its absence in other semimetals pushed beyond the quantum limit~\cite{Zhu2012}. While the one-dimensional spectrum is a generic feature of the three-dimensional electron gas confined to its lower Landau level, exciton formation is not. In most semimetals with heavy atoms,  the electric permittivity is large. Therefore, Coulomb attraction between holes and electrons is attenuated, hindering the formation of excitons. The electric permittivity in bismuth, for example, is twenty times larger than in graphite~(see the supplement for more discussions\cite{SM}). Another difference between graphite and bismuth  is the  evolution of mass and carrier density across the quantum limit. The unavoidable enhancement in DOS  due to Landau level degeneracy leads to an increase in carrier density in bismuth and an increased mass in graphite. As a consequence, the latter becomes a strongly correlated electron system at high magnetic fields.

One open question is the origin of the larger Nernst signal in the EI state in the vicinity of the critical point. Any quantitative analysis, however, requires a more complete set of data in order to quantify the magnitude of the transverse thermoelectric conductivity $\alpha_{xy}$, whose amplitude reflects the ratio of entropy to magnetic flux~\cite{Bergman,Behnia2016}. Availability of DC fields above the present ceiling of 45 T would lead to multiprobe studies of the critical point unveiled by the present study.

In summary, we carried out pulsed-field Nernst measurements in graphite up to 60 T. We found  a 47 T anomaly in the Nernst response and identified it as the result of the simultaneous evacuation of two Landau subbands, an electron-like and a hole-like one. The Nernst anomaly suddenly bifurcates to two distinct anomalies marking the boundaries of the field-induced state below 9.2 K. We showed that the BEC condensation temperature of excitons is expected to occur close to this temperature.

\matmethods{
\subsection*{Samples}
The Kish graphite samples we used in the experiment were obtained commercially. The summary of sample information are listed in the following table:
\begin{table}[!hbp]
\begin{center}
\begin{tabular}{|c|c|c|c|c|}

\hline
 Sample  & type & dimension (mm$^3$) \\
\hline
K1& kish &1$\times$0.95$\times$0.04\\
\hline
K2 & kish &0.95$\times$0.9$\times$0.02\\
\hline
K3 & kish &1.1$\times$1.2$\times$0.06\\
\hline
\end{tabular}
\caption{A description of the samples used in this study.}
\end{center}
\end{table}

\subsection*{The Nernst Measurement under Pulsed Field} The measurement of Nernst effect under pulsed field was performed in WHMFC in Wuhan. The signal was recorded by high-speed digitizer PXI-5922 made by National Instrument running at 2 MHz rate. More detailed information to obtain the authentic Nernst signal is provided in the SI Appendix.

\subsection*{Data  Availability} All data are made available within the article or SI Appendix.
}

\showmatmethods{} 

\acknow{This work was supported by the National Key Research and Development Program of China (Grant no. 2016YFA0401704), the National Science Foundation of China (Grant nos.51861135104 and 11574097) and Fundamental Research Funds for the Central Universities (Grant no. 2019kfyXMBZ071). In France, it was supported by the Agence Nationale de la Recherche  (ANR-18-CE92-0020-01; ANR-19-CE30-0014-04) and by Jeunes Equipes de l$'$Institut de Physique du Coll\`ege de France.Z.Z. acknowledges useful discussions with Ryuichi Shindou and Yuanchang Li.}

\showacknow{} 


\begin{thebibliography}{99}

\bibitem{Leggett} A. J. Leggett, Quantum liquids.\textit{ Science} \textbf{319}, 1203-1205 (2008).
 \bibitem{Anderson} M. H. Anderson, J. R. Ensher, M. R. Matthews, C. E. Wieman, E. A. Cornell, Observation of Bose-Einstein condensation in a dilute atomic vapor.\textit{ Science} \textbf{269}, 198-201 (1995).

\bibitem{Einstein} A. Einstein, Quantentheorie des einatomigen idealen Gases, (Sitzsungberichte der Preussische Akademie der Wissenschaften, Berlin, 1924),vol.3, pp.261-267.

\bibitem{Silvera}  F. I. Silvera, Bose-Einstein condensation. \textit{Am. J. Phys.} \textbf{65}, 570 (1997)
 \bibitem{Mott}  N. F. Mott, The transition to the metallic state.\textit{ Philos. Mag.} A \textbf{6}, 287-309 (1961).
 \bibitem{Snoke}  D. Snoke, Spontaneous Bose coherence of excitons and polaritons.\textit{ Science} \textbf{298}, 1368-1372 (2002).

 \bibitem{Eisenstein}  J. P. Eisenstein, A. H. MacDonald, Bose-Einstein condensation of excitons in
bilayer electron systems.\textit{ Nature} \textbf{432}, 691-694 (2004).
  \bibitem{Jerome1967}  D. J\'{e}rome, T. M. Rice, W. Kohn, Excitonic insulator.\textit{ Phys. Rev.} \textbf{158}, 462 (1967).
 \bibitem{Kogar}  A. Kogar \textit{et al.,}
 Signatures of exciton condensation in a transition metal dichalcogenide. \textit{Science} \textbf{358}, 1314-1317 (2017).
 \bibitem{Kono2017}Y. Lu \textit{et al.,} Zero-Gap semiconductor to excitonic insulator transition in Ta$_2$NiSe$_5$. \textit{Nat. Commun.} \textbf{8}, 14408 (2017).

 \bibitem{Butov2007}  L. V. Butov, Cold exciton gases in coupled quantum well structures.\textit{ J. Phys. Condens. Matter} \textbf{16}, 295202 (2007).
 \bibitem{Kim}  X. Liu, K. Watanabe, T. Taniguchi, B. I. Halperin, P. Kim,   Quantum Hall drag of exciton condensate in graphene.\textit{ Nat. Phys.} \textbf{13}, 746-750 (2017).
 \bibitem{Dean}  J. I. A. Li, T. Taniguchi, K. Watanabe, J. Hone, C. R. Dean, Excitonic superfluid phase in double bilayer graphene. \textit{ Nat. Phys.} \textbf{13}, 751-755 (2017).
 \bibitem{Burg} G. W. Burg \textit{et al.,}  Strongly enhanced tunneling at total charge neutrality in double-bilayer graphene-WSe$_2$ heterostructures. \textit{Phys. Rev. Lett.} \textbf{120}, 177702 (2018).
 \bibitem{Zefang}  Z. Wang \textit{et al.,}
 Evidence of high-temperature exciton condensation
 in two-dimensional atomic double layers. \textit{ Nature} \textbf{574}, 76-80 (2019).
 \bibitem{Karni}  O. Karni \textit{et al.,}
 Infrared interlayer exciton emission in MoS$_2$/WSe$_2$ heterostructures. \textit{Phys. Rev. Lett.} \textbf{123}, 247402 (2019).
\bibitem{Tanuma1981} S. Tanuma, R. Inada, A. Furukawa, O. Takahashi, Y. Iye, \textit{Physics in High Magnetic Fields,} S. Chikazumi, N. Miura, Eds. (Springer, Berlin, 1981), pp. 274-283.
\bibitem{Yaguchi2009} H. Yaguchi, J. Singleton, A high-magnetic-field-induced density-wave state in graphite. \textit{J. Phys. Condens. Matter.} \textbf{21}, 344207 (2009).
\bibitem{FauqueBook} B. Fauqu\'{e}, K. Behnia, \textit{ Basic Physics of Functionalized Graphite},  P. D. Esquinazi, Eds.  (Springer, 2016) Chap. 4.
\bibitem{Yoshioka} D. Yoshioka, H. Fukuyama, Electronic phase transition of graphite in a strong magnetic field. \textit{ J. Phys. Soc. Jpn.} \textbf{50}, 725-726 (1981).

\bibitem{Arnold2017}F. Arnold \textit{et al.,} Charge density waves in graphite: Towards the magnetic ultraquantum limit. \textit{Phys. Rev. Lett.} \textbf{119}, 136601 (2017).

\bibitem{Brandt1988} N. B. Brandt, S. M. Chudinov, Y. G. Ponomarev, \textit{ Modern Problems in Condensed Matter Sciences},  V. M. Agranovich, A. A. Maradudin, Eds. (North-Holland, Amsterdam, 1988), Vol. 20.

\bibitem{ZhuNernstGraphite} Z. Zhu, H. Yang, B. Fauqu\'{e}, Y. Kopelevich, K. Behnia, Nernst effect and dimensionality in the quantum limit. \textit{ Nat. Phys.} \textbf{6}, 26-29 (2010).
\bibitem{Schneider} J. M. Schneider, B. A. Piot, I. Sheikin,  D. K. Maude, Using the de Haas-van Alphen effect to map out the closed three-dimensional Fermi surface of natural graphite. \textit{Phys. Rev. Lett.} {\bf{108}}, 117401 (2012)
\bibitem{Benoit2013} B. Fauqu\'{e} \textit{et al.,} Two phase transitions induced by a magnetic field in graphite. \textit{ Phys. Rev. Lett.} \textbf{110}, 266601 (2013).
\bibitem{Zhu2019}Z. Zhu \textit{et al.,} Graphite in 90 T: Evidence for strong-coupling excitonic pairing. \textit{ Phys. Rev. X} \textbf{9}, 011058 (2019).
 \bibitem{ZhuEI2017} Z. Zhu \textit{et al.,} Magnetic field tuning of an excitonic insulator between the weak and strong coupling regimes in quantum limit graphite. \textit{ Sci. Rep.} \textbf{7}, 1733 (2017).
 \bibitem{Leboeuf2017}D. LeBoeuf \textit{et al.}, Thermodynamic signatures of the field-induced states of graphite. \textit{Nat. Commun.} \textbf{8}, 1337 (2017).
\bibitem{Takada1998} Y. Takada, H. Goto, Exchange and correlation effects in the three-dimensional electron gas in strong magnetic fields and application to graphite. \textit{J. Phys. Condens. Matter.} \textbf{10}, 11315-11325 (1998).
 \bibitem{Slonczewski1958}J. C. Slonczewski, P. R. Weiss. Band Structure of Graphite. \textit{Phys. Rev.} \textbf{109}, 272-279 (1958).
\bibitem{McClure1957}J. W. McClure. Band structure of Graphite and de Haas-van Alphen effect. \textit{ Phys. Rev.} \textbf{108}, 612-618 (1957).


\bibitem{Behnia2016} K. Behnia, H. Aubin, Nernst effect in metals and superconductors: A review of concepts and experiments. \textit{Rep. Prog. Phys.} \textbf{79}, 046502 (2016)

\bibitem{Zhu2011} Z. Zhu \textit{et al.,} Nernst quantum oscillations in bulk semi-metals. \textit{J. Phys. Condens. Matter} \textbf{23}, 094204 (2011)

 \bibitem{ProustNernst2009}J. Levallois, K. Behnia, J. Flouquet, P. Lejay, C. Proust. On the destruction of the hidden order in URu$_2$Si$_2$ by a strong magnetic field. \textit{Europhys. Lett} \textbf{85}, 27003 (2009).

 \bibitem{SM} Supplementary Material
 \bibitem{Benoit2011} B. Fauqu\'{e}, Z. Zhu, T. Murphy, K. Behnia, Nernst response of the Landau tubes in graphite across the quantum limit. \textit{Phys. Rev. Lett.} \textbf{106}, 246405 (2011).
 \bibitem{Akiba_2015}K. Akiba  \textit{et al.,} Possible excitonic phase of graphite in the quantum limit state. \textit{ J. Phys. Soc. Japan} \textbf{84}, 054709 (2015).
 \bibitem{QO2016} A. P. Morales \textit{et al.,} Thermoelectric power quantum oscillations in the ferromagnet UGe$_2$. \textit{Phys. Rev. B}  \textbf{93}, 155120 (2016).
\bibitem{QO2019} X. Xu \textit{et al.,}  Quantum oscillations in the noncentrosymmetric superconductor and topological nodal-line semimetal PbTaSe$_{2}$. \textit{Phys. Rev. B} \textbf{99}, 104516 (2019)

\bibitem{Marcenat} C. Marcenat \textit{et al.}, Wide critical
fluctuations of the field-induced phase transition in graphite. arXiv: 2011.02435.

\bibitem{London1938}F. London, The $\lambda$-phenomenon of liquid helium and the Bose-Einstein degeneracy. \textit{Nature} \textbf{141},643(1938).

\bibitem{Vakarchuk} I. Vakarchuk, O. Hryhorchak, V. Pastukhov, and R. Prytula,  Effective mass of $^4$He atom in superfluid and normal phases. Ukrainian J. Phys., \textbf{61}, 29 (2019).

\bibitem{Zhu2017} Z. Zhu \textit{et al.}, Emptying Dirac valleys in bismuth using high magnetic fields. \textit{Nat. Commun.} \textbf{8}, 15297 (2017).

\bibitem{Kozlov1965}A. N. Kozlov, L. A. Maksimov,  The metal-dielectric divalent crystal phase transition.  \textit{Zh. Eksperim. i Teor. Fiz.} {\textbf{48}}, 1184 (1965) [English translation: \textit{Soviet Phys.-JETP} {\textbf{21}}, 790 (1965)].
\bibitem{Iye1982}Y. Iye \textit{et al.,} High-magnetic-field electronic phase transition in graphite observed by magnetoresistance anomaly. \textit{ Phys. Rev. B} \textbf{25}, 5478-5485 (1982).

\bibitem{Zhu2012} Z. Zhu \textit{et al.},  Landau spectrum and twin boundaries of bismuth in the extreme quantum limit.
\textit{Proc. Natl. Acad. Sci. U.S.A.} \textbf{109},  14813-14818 (2012)


\bibitem{Bergman} D. L. Bergman, V. Oganesyan, Theory of dissipationless Nernst Effects. \textit{ Phys. Rev. Lett.} \textbf{104}, 066601 (2010).
\end{thebibliography}

\begin{thebibliography}{99}
\bibitem{Yaguchi2009_SM} H. Yaguchi, J. Singleton, A high-magnetic-field-induced density-wave state in graphite. \textit{J. Phys. Condens. Matter.} \textbf{21}, 344207 (2009).
\bibitem{ZhuNernstGraphite_SM} Z. Zhu, H. Yang, B. Fauqu\'{e}, Y. Kopelevich, K. Behnia, Nernst effect and dimensionality in the quantum limit. \textit{ Nat. Phys.} \textbf{6}, 26-29 (2010).

 \bibitem{Mott-SM}  N. F. Mott, The Transition to the Metallic state. \textit{ Philos. Mag.} \textbf{6}, 287-309 (1961).
\bibitem{Knox1963-SM} R. S. Knox, \textit{Solid State Physics},( edited by F. Seitz and D. Turnbull Academic Press Inc. , New York, 1963), Suppl. 5, p. 100.

\bibitem{Jerome1967-SM}  D. J\'{e}rome, T. M. Rice, W. Kohn, Excitonic insulator. \textit{ Phys. Rev.} \textbf{158}, 462 (1967).
\bibitem{Halperin1968-SM}B. I. Halperin, T. M. Rice, Possible anomalies at a semimetal-semiconductor transition.  \textit{Rev. Mod. Phys.} \textbf{40},755-766 (1968).

\end{thebibliography}

\newpage

\renewcommand{\thesection}{S\arabic{section}}
\renewcommand{\thetable}{S\arabic{table}}
\renewcommand{\thefigure}{S\arabic{figure}}
\renewcommand{\theequation}{S\arabic{equation}}

\setcounter{section}{0}
\setcounter{figure}{0}
\setcounter{table}{0}
\setcounter{equation}{0}

{\large\bf Supplemental Material for ``Critical point for Bose-Einstein condensation of excitons in graphite''}

\section{The Nernst Measurement under Pulsed Field}

The Nernst effect under pulsed field was performed in WHMFC in Wuhan. The signal was recorded by high-speed digitizer PXI-5922 made by National Instrument running at 2 MHz rate.

\begin{figure}[h]
\includegraphics[width=9cm]{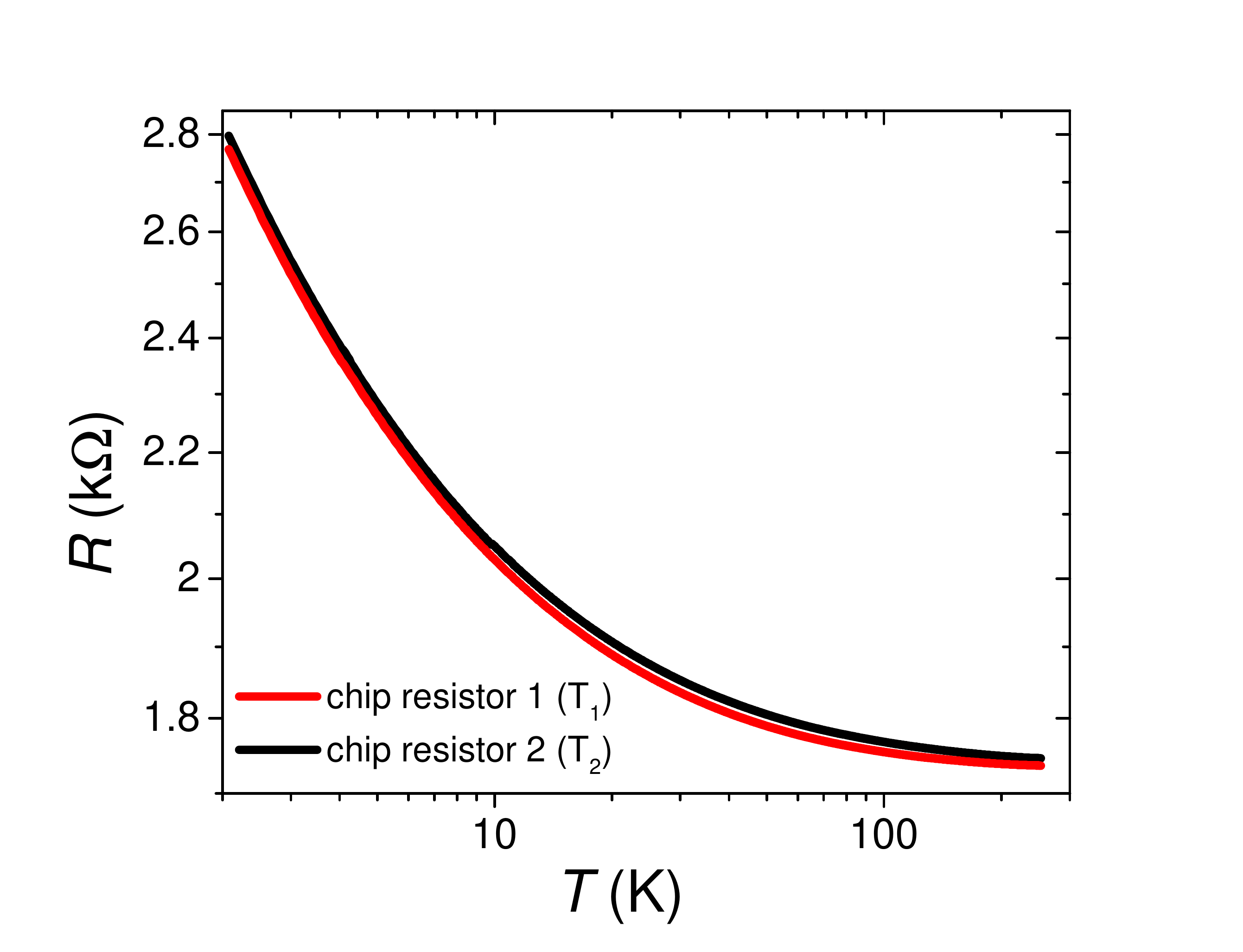}
\caption{ The R-T curves of the two chip resistors used as temperature calibration in the Nernst measurements for K3. }.
\label{Sfig:RT}
\end{figure}

\begin{figure}[t]
\includegraphics[width=9cm]{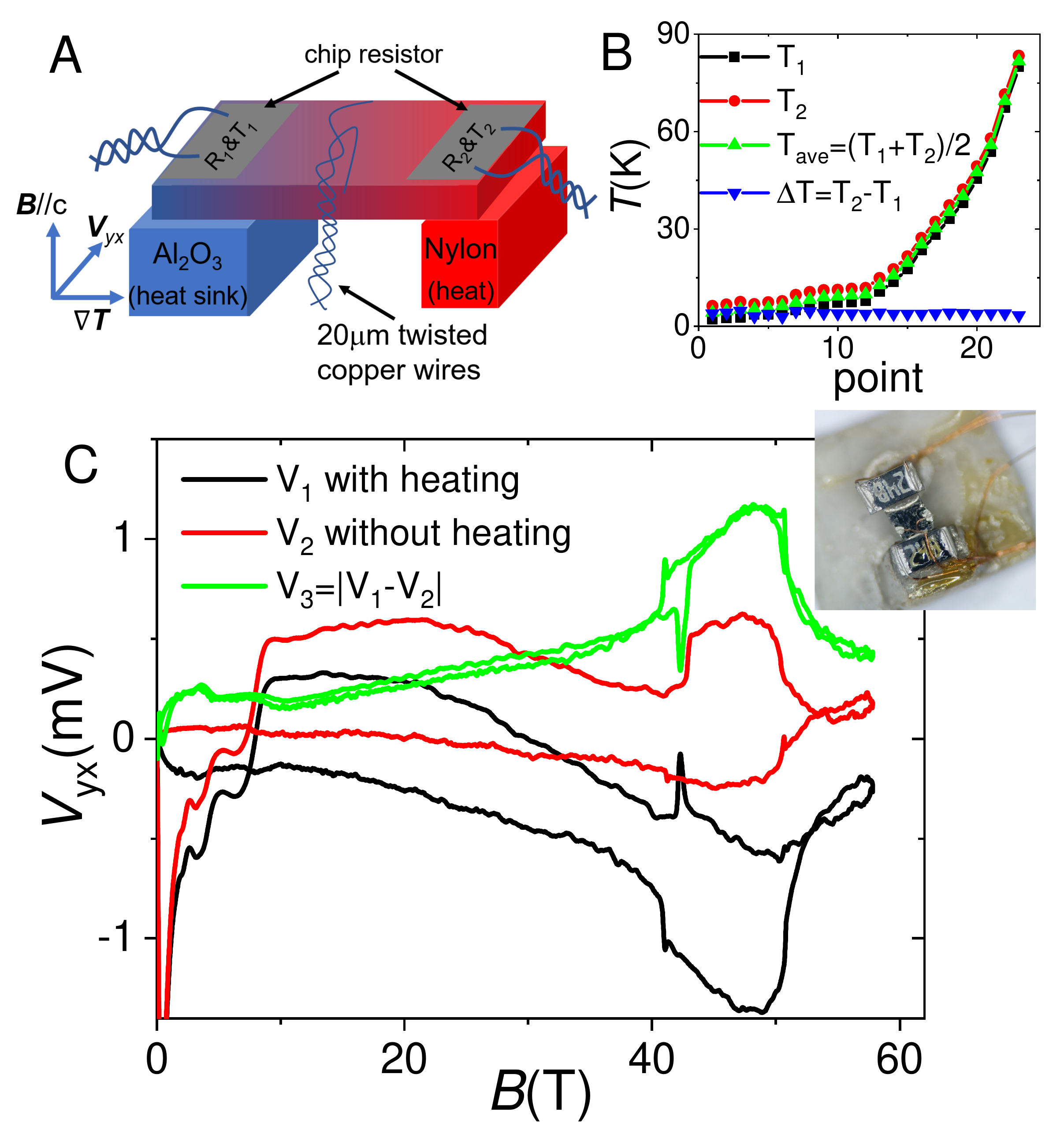}
\caption{\textbf{To obtain the authentic Nernst signal under pulsed field:} ({\textit{A}}) The sketch of the Nernst effect setup under pulsed field. The temperature gradient was established by applied current to the chip resistor on the Nylon side. ({\textit{B}}) The estimated temperature difference and average temperature from the two calibrated resistors after applied heating current. ({\textit{C}}) A typical result of Nernst data obtained at 7.3 K with a photo of the setup inserted. The authentic Nernst  signal (green) is obtained by subtracting the signal with heating (black) from the signal without heating (red).}
\label{SNernst}
\end{figure}

The Fig. \ref{Sfig:RT} shows the temperature dependence of resistance of the two chip resistors used to measure temperatures and temperature differences. Before measuring the calibration curves, the chip resistors were worn as thin as possible and trained several times in liquid nitrogen to stabilize the chip resistors. The resistors then were cooled down and calibrated to a known thermometer.

To produce a large thermal gradient and reduce the eddy current effect, the samples were suspended on two types of insulator supports (See Fig. \ref{SNernst}{\textit{A}}). The heat sink was made of a good thermal conductor: sapphire (Al$_2$O$_3$) and the heat source side was made of a bad thermal conductor: Nylon. Two RuO$_2$-type chip resistors $R_1$ and $R_2$ were mounted on two sides of sample with silver paint to measure $T_1$ (the heat sink side) and $T_2$ (the heat source side), which are served as heater also thermometer. Before the Nernst effect measurement, the resistances of the resistors were calibrated up to 250 K (See Fig. \ref{Sfig:RT}). During the measurements, the whole system was regulated at desired temperatures.  To prevail background signal, the temperature differences had to be large around 4 K in whole measurements. The estimated sample temperature was calculated by $T_{ave}=(T_1+T_2)/2$ shown in the Fig. \ref{SNernst}{\textit{B}}. However, the actual temperature below 9.2 K was calibrated by the transition field of $\alpha$\cite{Yaguchi2009_SM} without heating during a pulsed field and it is proved to be quite accurate from the sharpness of the rising at $\alpha$ and falling at $\alpha'$ instead of broad shapes. This temperature is close to the estimated temperature from thermometers on the setup. To reduce the loop signal caused by large $dB/dt$, 20 $\mu$m diameter twisted enameled copper wires were connected to transverse laterals of sample to measure Nernst signal. The photograph in Fig. \ref{SNernst}{\textit{C}} shows the actual setup. An in-plane thermal gradient was established before applying pulsed field along the c-axis of sample.
\begin{figure}
\includegraphics[width=9cm]{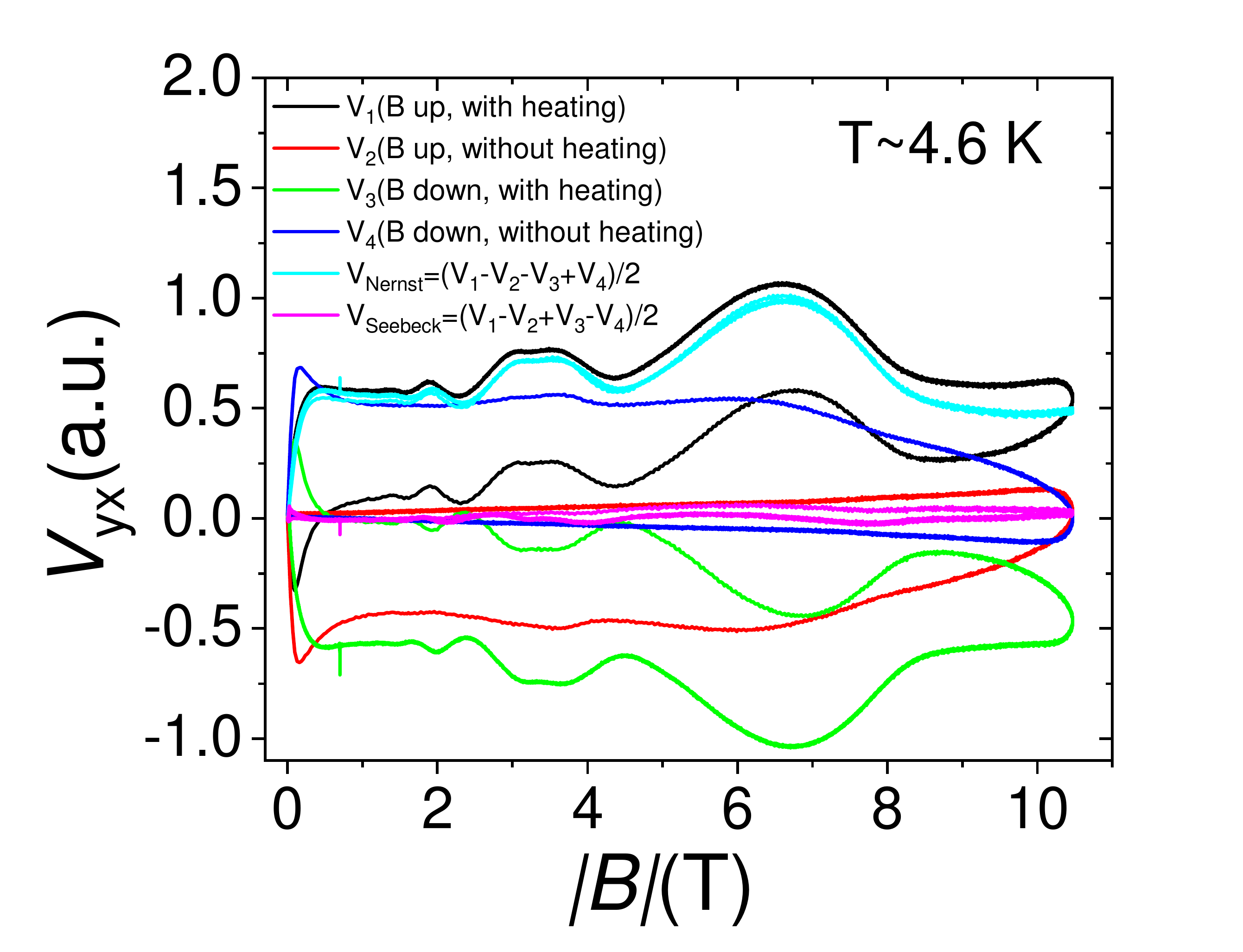}
\caption{ A typical result up to 10 T.  $V_{Nernst}$  (in cyan) is much larger than $V_{Seebeck}$ (in magenta). }.
\label{Sfig:overlap}
\end{figure}

\begin{figure}
\includegraphics[width=9cm]{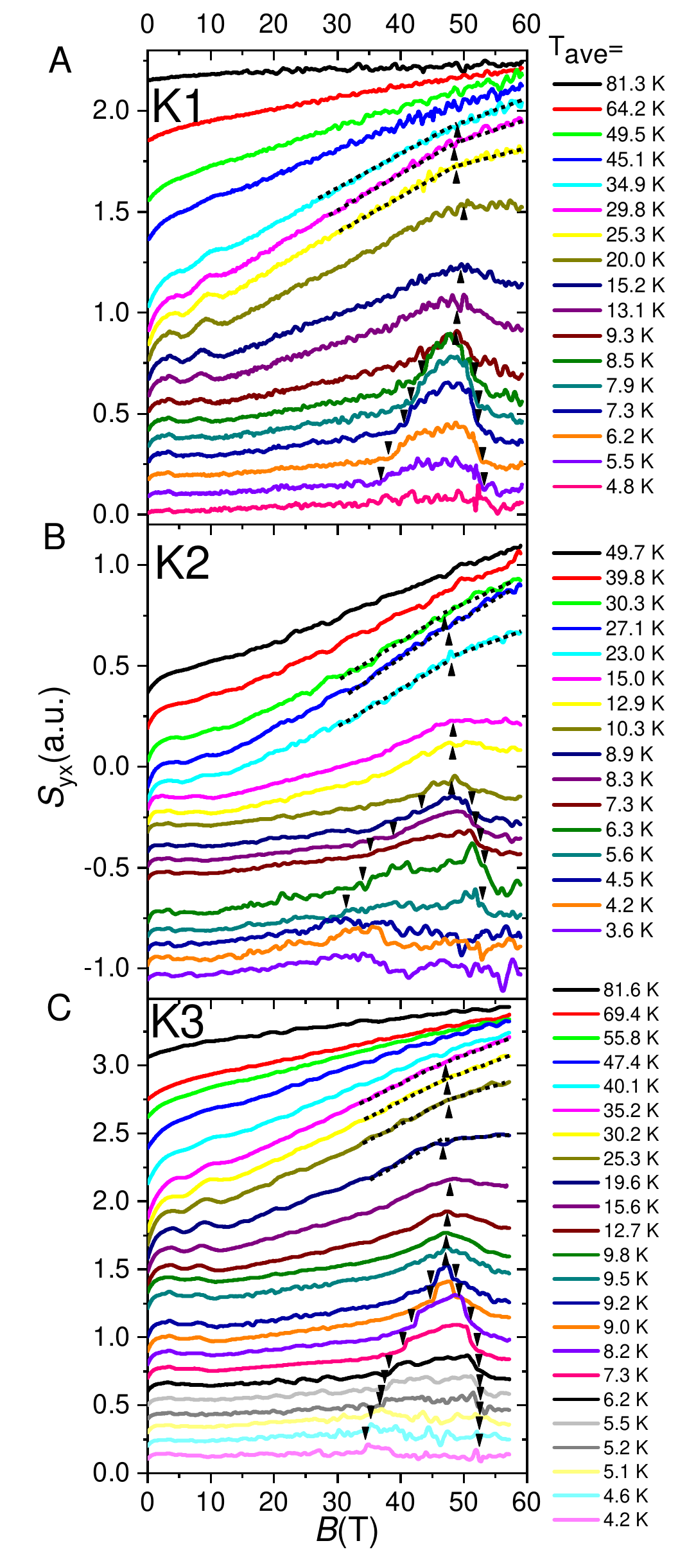}
\caption {({\textit{A}})-({\textit{C}}) Shifted Nernst data at different  temperature for K1, K2 and K3. The dash lines are to identify the 47 T peaks. Up-arrows indicate the 47 T peaks and down-arrows indicate $\alpha$ as well as $\alpha'$ transitions. The data are consistent in the three samples.}
\label{fig:SNernst}
\end{figure}

Fig. \ref{SNernst}{\textit{C}} shows a typical Nernst measurement result at 7.3 K. The authentic Nernst signal (green) is obtained by subtracting the signal with heating (black) from the signal without heating (red). The measured signal in pulsed field includes three parts: the authentic Nernst signal $V_{yx}$, the varying-field induced voltage (electromagnetic induction) and the Seebeck signal from a misalignment of potential contacts. We obtained the authentic Nernst signal by a two-round measurement: with and without heating sample during pulsed fields. With temperature gradient under pulsed field, the voltage signal can be expressed as $V_1=V_{yx}\pm ({A} \rm{d}B_1/\rm{d}t+I_{eddy}R)\pm V_{xx}$ (the black curve in Fig. \ref{SNernst}{\textit{C}}).  $A$ is the wire loop area perpendicular to the field, which can be reduced or cancelled by using twisted wires. $I_{\rm{eddy}}$ is the eddy currents (Foucault's currents) and proportional to the field variation rate $\rm{d}B/\rm{d}t$, $R$ is the resistance along the eddy current and $V_{xx}$ is the Seebeck signal. Since the Seebeck effect is around one-order smaller comparing to Nernst effect in graphite\cite{ZhuNernstGraphite_SM}, we can neglect this Seebeck part. Indeed, the Nernst signal is almost same by reversing the orientation of the pulsed magnetic field. Meanwhile, without temperature gradient under pulsed field, we can obtain only the second part: $V_2=\pm(A \rm{d}B_2/\rm{d}t+I_{eddy}R)$ (the red curve in Fig. \ref{SNernst}{\textit{C}}). $\rm{d}B_1/\rm{d}t$ and $\rm{d}B_2/\rm{d}t$ can be easily to make same if we pulse the magnet to a same field at a same condition for two pulses. Finally, the Nernst signal is deduced by $V_{yx} \simeq V_3=|V_1-V_2|$ which is the green curve in Fig. \ref{SNernst}{\textit{C}}. This well overlap of authentic Nernst part as the field is rising and falling indicates the validation of measurement and also the sample was not heated much during the pulses.

\section{Reproducibility of the measurements}

A typical result up to 10 T of reversing the orientation of magnetic field is shown in Fig. \ref{Sfig:overlap}. After symmetrization/asymmetrization, the $V_{Nernst}$ is well-overlapped and much larger than $V_{Seebeck}$. So we can obtain the Nernst signal by two shoots( with and without heating) in the same direction .

The whole Nernst signal data of K3 are shown in Fig. \ref{fig:SNernst}{\textit{C}}. We also measured Nernst effect in different Kish graphite samples named K1 and K2. We obtained similar results in the two samples shown in Fig. \ref{fig:SNernst}{\textit{A}} and \ref{fig:SNernst}{\textit{B}}. All data have been shifted for clarity. The 47 T peaks can be observed clearly in all three samples. Disappearance of the peak is also similar at 35 K and the critical temperature of phase A is as well as consistent in the three samples.

\section{The EI scenario in other semi-metals}

The concept of an excitonic insulator (EI) was proposed by Mott~\cite{Mott-SM} in the context of semimetals. It was extended to semiconductors by Knox~\cite{Knox1963-SM}. This led to a search for the EI in semimetals. Semi-metallic elements Bi, Sb, and As appeared promising at the first sight. However, the large electric permittivity in these semimetals hinders the formation of EI.

 The binding energy of an exciton is $E_{\rm{B}}= (\mu/m_0)(1/{\epsilon}^2)R_{\rm{y}}$ \cite{Jerome1967-SM,Halperin1968-SM}.  Here $\mu$, $m_0$, $\epsilon$ and $R_{\rm{y}}$ are the reduced mass, the free electron mass, the dielectric constant and the Rydberg energy. With a typical mass of $\mu=0.01 m_0$ and a dielectric constant of $\epsilon=100$ for these semimetals, the binding energy would be $E_{\rm{B}}=1.36\times10^{-5}$eV$~\sim~0.1$ K

\end{document}